# Reciprocal Space Mapping for Dummies

Samuele Lilliu[1], Thomas Dane[2]

[1]University of Sheffield, Hicks Building, Hounsfield Road, Sheffield, S3 7RH (UK), [2]European Synchrotron Radiation Facility, BP 220, Grenoble F-38043, France.

## Abstract

Grazing Incidence X-ray Diffraction (GIXD) is a surface sensitive X-ray investigation technique (or geometry configuration) that can reveal the structural properties of a film deposited on a flat substrate[1-4]. The term grazing indicates that the angle between the incident beam and the film is small (typically below 0.5°). This essential technique has been employed on liquid crystals[5,6], nanoparticles and colloids[7,8], nanostructures[9], corrosion processes[10,11], polymers[12-14], bio-materials[15-17], interfaces[11,18], materials for solar cells[19-25], photodiodes[26], and transistors[27,28], etc. Diffraction patterns in GIXD geometry are typically captured with a 2D detector, which outputs images in pixel coordinates. A step required to perform analyses such as grain size estimation, disorder, preferred orientation, quantitative phase analysis of the probed film surface, etc.[29], consists in converting the diffraction image from pixel coordinates to the momentum transfer or scattering vector in sample coordinates (the 'reciprocal space mapping')[30]. This momentum transfer embeds information on the crystal or polycrystal and its intrinsic rotation with respect to the substrate. In this work we derive, in a rigorous way, the reciprocal space mapping equations for a '3D+1S' diffractometer in a way that is understandable to anyone with basic notions of linear algebra, geometry, and X-ray diffraction.

## Introduction

Two-dimensional X-ray diffraction (XRD$^2$) is a well-established technique in the field of X-ray diffraction (XRD)[29]. The main complication of data analysis in XRD$^2$ compared to one dimension XRD is the remapping of detector pixel coordinates into momentum transfer or scattering vector coordinates. The geometry of a two-dimensional X-ray diffraction system consists of three reference systems: laboratory, detector, and sample. The laboratory coordinate system is the reference or global three dimensional coordinate system. The sample coordinate system shares the same origin with the laboratory system but its basis can be oriented in a different way. The detector coordinates system is two-dimensional. The three reference systems are represented by orthogonal Cartesian basis[31-34].

In this work we employ a '3D+1S' diffractometer, where three diffractometer circles (3D) are used for moving the 2D detector across the Ewald sphere, and one circle is used for orienting the sample. The simplest detector orientation corresponds to the situation in which the normal to the detector is coincident to the X-ray incident beam or, equivalently, points towards the origin of the laboratory coordinate system, where the sample is placed (detector circle angles are zero). The simplest sample orientation corresponds to the grazing case in which the sample surface is parallel to the direct beam (sample circle angle is zero). Let's assume that our sample is a powder of randomly oriented crystals and consider a diffracting plane. The diffracting plane would project a cone of rays (Debye-Scherrer ring) onto the detector[1]. Since the detector is perpendicular to incident beam and the sample is parallel to the incident beam, the projection of the diffracting cone onto the detector would be a perfect circle. In this case, converting detector's pixel coordinates into scattering vector coordinates is non-trivial. However, one might want to move the detector to non-zero azimuth and elevation to visualize extra Debye-Scherrer rings. This complicates things and introduces distortions in the diffraction pattern. In the most general case the Debye-Scherrer rings will have a tilted ellipse shape, which result from the intersection of the diffracting cone with the tilted detector plane. In this case, the conversion between pixels and scattering vector coordinates is more complicated.

We start our discussion by introducing the diffractometer geometry and the laboratory coordinate system. We then show how to construct the rotation matrices for the detector rotation (3D). The next step is to project a ray with the same direction of the exit wave versor $\hat{k}_f$ onto a rotated detector. The projected ray is converted into the two-dimensional detector coordinates. At this point we have a set of equations that convert the exit wave versor ray into an image point. These equations can be inverted so that, given an image point, we can retrieve the ray information. We then introduce the effect of the rotation of one sample circle and derive the equations required to reconstruct the scattering vector in sample coordinates, which contains information on the crystal and its intrinsic rotation with respect to the substrate.

### Diffractometer Geometry and Laboratory System

For simplicity we refer to a common synchrotron beamline setup, such as the one employed at the XMaS beamline (BM28, the European Synchrotron, ESRF)[35], even though the following discussion can be applied to similar configurations. As shown in **Figure 1a-b,** a chamber with a sample holder is mounted on a 12 circle Huber diffractometer, placing the sample at the origin of the laboratory and sample coordinate system.

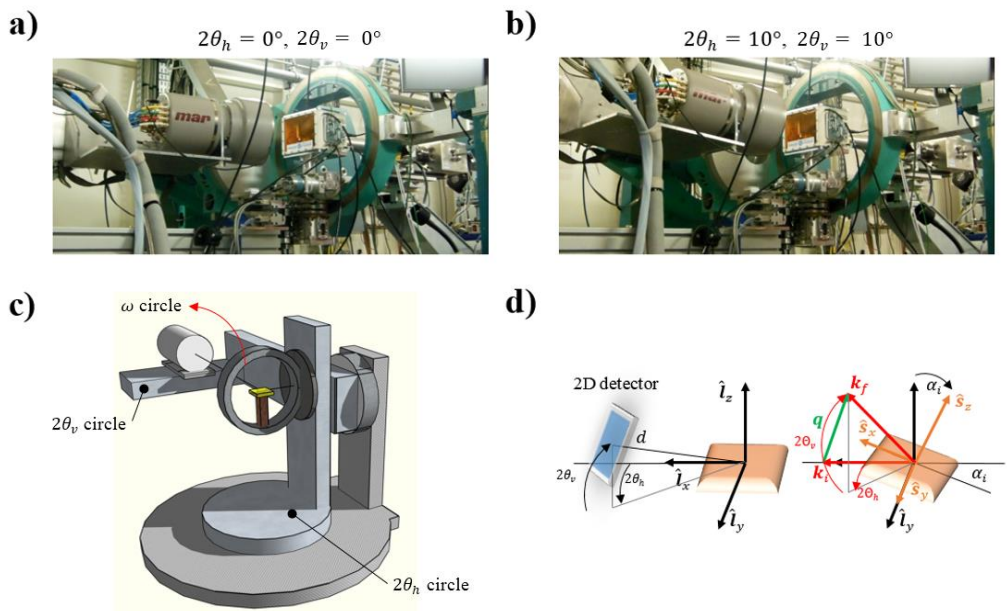

**Figure 1 - Diffractometer Geometry**. **a)** Detector at the 'origin' position with $2\theta_h = 0°$ and $2\theta_v = 0°$. **b)** Detector at $2\theta_h = 10°$ and $2\theta_v = 10°$. **c)** Diffractometer sketch. **d)** Laboratory $(\hat{l}_x, \hat{l}_y, \hat{l}_z)$ and sample $(\hat{s}_x, \hat{s}_y, \hat{s}_z)$ coordinate systems; with the blue rectangle representing the detector ($d$ is the sample-detector distance, $2\theta_v$ and $2\theta_h$ are the detector angular coordinates, $\omega$ is an extra motor that spins the detector about its normal) and the orange rectangle representing the sample. The sample in the second picture of d) is tilted of an out-of-plane incident angle $\alpha_i$. The incident wave vector (direct X-ray beam) is $k_i$ and the diffracted wave vector is $k_f$, with $2\Theta_h$ and $2\Theta_v$ being its angular coordinates. The scattering vector is $q = k_f - k_i$. Wave vectors and scattering vector are discussed later in the text.

Here we consider two detector circles ($2\theta_h$ and $2\theta_v$) and one sample pitch circle ($\alpha_i$). We also add an extra degree of freedom for the detector though a third detector circle $\omega$, which rotates the detector about its normal. The diffractometer geometry is depicted in **Figure 1d**.

The global coordinate system (or laboratory) coordinate (or reference) system is the standard orthogonal Cartesian basis with origin at the diffractometer centre $O$:

$$\mathrm{L} = [\hat{\boldsymbol{l}}_x \quad \hat{\boldsymbol{l}}_y \quad \hat{\boldsymbol{l}}_z] = \mathrm{I} = \begin{bmatrix} 1 & 0 & 0 \\ 0 & 1 & 0 \\ 0 & 0 & 1 \end{bmatrix} \quad (1)$$

where $\hat{\boldsymbol{l}}_x, \hat{\boldsymbol{l}}_y, \hat{\boldsymbol{l}}_z$ are unit vectors (versors), and I is the identity matrix[31-34]. The horizontal detector circle moves the detector counterclockwise with respect to $\hat{\boldsymbol{l}}_z$ for $2\theta_h > 0°$ (see **Figure 1d**). The vertical detector circle moves the detector clockwise (upwards) with respect to the $\hat{\boldsymbol{l}}_y$ axis for $2\theta_v > 0°$ (when $2\theta_h = 0°$). An out-of-plane incidence angle or pitch $\alpha_i > 0°$ moves the sample clockwise with respect to $\hat{\boldsymbol{l}}_y$. The extra detector circle $\omega$ is not installed at the XMaS beamline, but could be used to rotate counterclockwise the detector about its normal allowing combined grazing incidence wide angle x-ray scattering (GIWAXS) and grazing incidence small angle x-ray scattering (GISAXS) for example[36].

Tracking the Detector Movement
For simplicity we assume that the detector has a square shape. Given a fixed point $P$ on the detector (e.g. its top-right corner), we would like to know the location of this point in the laboratory system when the $2\theta_h$, $2\theta_v$, and $\omega$ circle angles are non-zero. This can be done by constructing a rotation matrix that will rotate the point $P$. When constructing the rotation matrix we need to pay attention to the way the circles are mounted. In our case the $\omega$ circle is mounted on the $2\theta_v$ circle, which is mounted on the $2\theta_h$ circle (**Figure 1c**). The way to build the total rotation matrix proceeds from the inner circle to the outer circle as $\mathrm{R} = \mathrm{R}_{out}\mathrm{R}_{out-1} \cdots \mathrm{R}_{in+1}\mathrm{R}_{in}$ [30]. The following calculations can be performed in MATLAB with **Script 2**, Supplementary Information.

The first rotation matrix $\mathrm{R}_x$ is counterclockwise[31-34] with respect to $\hat{\boldsymbol{l}}_x = [1 \quad 0 \quad 0]^\mathrm{T}$ and represents the $\omega$ detector rotation about its normal:

$$\mathrm{R}_x = \begin{bmatrix} 1 & 0 & 0 \\ 0 & \cos\omega & -\sin\omega \\ 0 & \sin\omega & \cos\omega \end{bmatrix} \quad (2)$$

The second rotation matrix $\mathrm{R}_y$ is counterclockwise about $-\hat{\boldsymbol{l}}_y = [0 \quad -1 \quad 0]^\mathrm{T}$ (or clockwise about $\hat{\boldsymbol{l}}_y$) and represents the $2\theta_v$ detector rotation in the vertical direction (elevation):

$$\mathrm{R}_y = \begin{bmatrix} \cos 2\theta_v & 0 & -\sin 2\theta_v \\ 0 & 1 & 0 \\ \sin 2\theta_v & 0 & \cos 2\theta_v \end{bmatrix} \quad (3)$$

Note again that the convention used here is that a positive $2\theta_v$ rotation is clockwise about $\hat{\boldsymbol{l}}_y$, while in ref. [36] a positive $2\theta_v$ rotation corresponds to a counterclockwise rotation about $\hat{\boldsymbol{l}}_y$.

The last rotation matrix $\mathrm{R}_z$ is counterclockwise about $\hat{\boldsymbol{l}}_z = [0 \quad 0 \quad 1]^\mathrm{T}$ and represents the $2\theta_h$ detector rotation in the horizontal direction (azimuth):

$$\mathrm{R}_z = \begin{bmatrix} \cos 2\theta_h & -\sin 2\theta_h & 0 \\ \sin 2\theta_h & \cos 2\theta_h & 0 \\ 0 & 0 & 1 \end{bmatrix} \quad (4)$$

The total rotation matrix is:

$$R_{xyz} = R_z R_y R_x$$
$$= \begin{bmatrix} \cos 2\theta_h \cos 2\theta_v & -\sin 2\theta_h \cos \omega - \cos 2\theta_h \sin 2\theta_v \sin \omega & \sin 2\theta_h \sin \omega - \cos 2\theta_h \sin 2\theta_v \cos \omega \\ \sin 2\theta_h \cos 2\theta_v & \cos 2\theta_h \cos \omega - \sin 2\theta_h \sin 2\theta_v \sin \omega & -\cos 2\theta_2 \sin \omega - \sin 2\theta_h \sin 2\theta_v \cos \omega \\ \sin 2\theta_v & \cos 2\theta_v \sin \omega & \cos 2\theta_v \cos \omega \end{bmatrix} \quad (5)$$

With this rotation, the standard basis versors $\hat{l}_x, \hat{l}_y, \hat{l}_z$ rotate and the rotated basis becomes $L' = R_{xyz}L = R_{xyz}I = R_{xyz}$. The rotated point $P'$, expressed in the L basis (and not in the L' basis) is located at:

$$P' = R_{xyz} P \quad (6)$$

Note that $R_z, R_y, R_x$ do not commute[37]. The matrix $R_{xyz}$ represents an orthogonal ($R_{z,y1}^{-1} = R_{z,y1}^T$) and rigid transformation, which preserves the orientation of the transformed vectors ($\det R_{z,y1} = 1$).
When $\omega = 0$ eq. (5) reduces to:

$$R_{yz} = R_z R_y = \begin{bmatrix} \cos 2\theta_h \cos 2\theta_v & -\sin 2\theta_h & -\cos 2\theta_h \sin 2\theta_v \\ \sin 2\theta_h \cos 2\theta_v & \cos 2\theta_h & -\sin 2\theta_h \sin 2\theta_v \\ \sin 2\theta_v & 0 & \cos 2\theta_v \end{bmatrix} \quad (7)$$

## Projecting the Exit Wave Vector onto the Detector

We define $k_i$ as the incident wave vector from the X-ray direct beam and $k_f$ as the scattered or diffracted (exit) wave vector (**Figure 1d**). The type of scattering considered here is due to the absorption of incident radiation with subsequent re-emission (elastic scattering) $|k_f| = |k_i| = k = 2\pi/\lambda$ where $\lambda$ is the wavelength. Generally speaking, when the direct beam hits a crystalline sample, a diffracting ray is irradiated from the sample itself. The ray direction is given by the direction of the exit wave versor $\hat{k}_f = k_f/k$. The angle between the incident and the exit vector is the Bragg angle $2\Theta$ and their difference is the scattering vector $q = k_f - k_i$. The exit wave vector describes a sphere of radius $2\pi/\lambda$ called the Ewald sphere. More details on X-ray diffraction can be found elsewhere [1,4,38,39].

Using the subscript $L$ to highlight that the following vectors are expressed in laboratory coordinates[30], we express the exit wave vector as:

$$\boldsymbol{k}_{fL} = k\hat{\boldsymbol{k}}_{fL} = k \begin{bmatrix} \cos 2\Theta_{vL} \cos 2\Theta_{hL} \\ \cos 2\Theta_{vL} \sin 2\Theta_{hL} \\ \sin 2\Theta_{vL} \end{bmatrix} = k \begin{bmatrix} k_{fLx} \\ k_{fLy} \\ k_{fLz} \end{bmatrix} \quad (8)$$

where the third expression in eq. (8) is in spherical coordinates and the last in Cartesian coordinates. The way azimuth $2\Theta_{hL}$ (angle between the projection of $\hat{k}_{fL}$ on the xy-plane and the x-axis) and elevation $2\Theta_{vL}$ (angle between $\hat{k}_{fL}$ and the xy-plane) are measured corresponds to the way azimuth and elevation are measured for the detector (see **Figure 1d**).

The incident wave vector in laboratory coordinates[30] points towards the $\hat{l}_x$ axis:

$$\boldsymbol{k}_{iL} = k\hat{\boldsymbol{k}}_{iL} = k \begin{bmatrix} 1 \\ 0 \\ 0 \end{bmatrix} \quad (9)$$

therefore the scattering vector in laboratory coordinates[30] is:

$$\boldsymbol{q}_L = \boldsymbol{k}_{fL} - \boldsymbol{k}_{iL} = k \begin{bmatrix} \cos 2\Theta_{vL} \cos 2\Theta_{hL} - 1 \\ \cos 2\Theta_{vL} \sin 2\Theta_{hL} \\ \sin 2\Theta_{vL} \end{bmatrix} \quad (10)$$

The angle between the incident and exit vector is $2\Theta_L = \angle \boldsymbol{k}_{fL}, \boldsymbol{k}_{iL} = \mathrm{acos}(\boldsymbol{k}_{fL} \cdot \boldsymbol{k}_{iL})/k^2 = \mathrm{acos}(\cos 2\Theta_{hL} \cos 2\Theta_{vL})$. It is easy to verify that $|\boldsymbol{q}| = (4\pi/\lambda) \sin \Theta_L$.

Let's consider a ray described by the product of a scalar $v = [0, \infty]$ (distance) and the exit wave versor $\widehat{\boldsymbol{k}}_{fL}$ (direction) pointing towards the unrotated detector and with its origin in the origin of the laboratory system:

$$P = v \widehat{\boldsymbol{k}}_{fL} \quad (11)$$

The length of the ray is controlled by the scalar $v$. If the detector circle angles are zero, the detector's normal (versor) is:

$$\widehat{\boldsymbol{n}} = \begin{bmatrix} 1 \\ 0 \\ 0 \end{bmatrix} \quad (12)$$

We would like to calculate $v$ so that $P$ is on the detector plane. The plane can be described by the scalar product[31-34]:

$$P \cdot \widehat{\boldsymbol{n}} = d \quad (13)$$

where $P$ is a point on the plane, and $d$ is the distance between the plane and the origin of the laboratory coordinates. Substituting eq. (11) in eq. (13) we get:

$$v = \frac{d}{\widehat{\boldsymbol{k}}_{fL} \cdot \widehat{\boldsymbol{n}}} \quad (14)$$

When the detector is at non-zero circle angles ($2\theta_h \neq 0$, $2\theta_v \neq 0$, and $\omega \neq 0$) $\widehat{\boldsymbol{n}}$ rotates according to (first $\mathrm{R}_{xyz}$ column):

$$\widehat{\boldsymbol{n}}' = \mathrm{R}_{xyz} \widehat{\boldsymbol{n}} = \begin{bmatrix} \cos 2\theta_h \cos 2\theta_v \\ \sin 2\theta_h \cos 2\theta_v \\ \sin 2\theta_v \end{bmatrix} \quad (15)$$

Therefore the ray intersects the detector at:

$$P' = v \widehat{\boldsymbol{k}}_{fL} = \frac{d \widehat{\boldsymbol{k}}_{fL}}{\widehat{\boldsymbol{k}}_{fL} \cdot \widehat{\boldsymbol{n}}'} = \frac{d}{\Delta} \begin{bmatrix} \cos 2\Theta_{vL} \cos 2\Theta_{hL} \\ \cos 2\Theta_{vL} \sin 2\Theta_{hL} \\ \sin 2\Theta_{vL} \end{bmatrix} \quad (16)$$

$\Delta = \cos 2\theta_h \cos 2\theta_v \cos 2\Theta_v \cos 2\Theta_h + \sin 2\theta_h \cos 2\theta_v \cos 2\Theta_v \sin 2\Theta_h + \sin 2\theta_v \sin 2\Theta_v$

If we use Cartesian coordinates for $\widehat{\boldsymbol{k}}_{fL}$:

$$\widehat{\boldsymbol{k}}_{fL} = \begin{bmatrix} k_{fLx} \\ k_{fLy} \\ k_{fLz} \end{bmatrix}$$

$$P' = v\widehat{\boldsymbol{k}}_{fL} = \frac{d\widehat{\boldsymbol{k}}_{fL}}{\widehat{\boldsymbol{k}}_{fL} \cdot \widehat{\boldsymbol{n}}'} = \frac{d}{\Delta}\begin{bmatrix} k_{fLx} \\ k_{fLy} \\ k_{fLz} \end{bmatrix} \quad (17)$$

$$\Delta = k_{fLx}\cos 2\theta_h \cos 2\theta_v + k_{fLy}\cos 2\theta_v \sin 2\theta_h + k_{fLz}\sin 2\theta_v$$

We observe that since $\widehat{\boldsymbol{k}}_{fL}$ is a versor $k_{fLx}^2 + k_{fLy}^2 + k_{fLz}^2 = 1$, and that therefore it only requires two elements to be described. With eq. (16) or eq. (17) we can calculate the intersection between a ray $v\widehat{\boldsymbol{k}}_{fL}$ with origin in the origin of the laboratory system and the plane passing by the detector. The intersecting point is represented in laboratory coordinates.

We will now represent the ray intersecting the detector plane in detector coordinates. The following calculations can be performed in MATLAB with **Script 4**, Supplementary Information. The image coordinates correspond to the way the image matrix is indexed. Depending on the convention used, the row index $i = 1$ (y-axis) and the column index $j = 1$ (x-axis) might correspond to the top left corner. If the image size is $\Delta y \times \Delta x$, then $i = \Delta y$ and $j = \Delta x$, corresponds to the bottom right corner. We would like to use a different coordinate system for the detector, having as the centre the pixel hit by the direct beam (incident wave vector). If the direct beam points towards the centre of the detector, we chose the centre of the detector reference system as $j_0 = \Delta x/2$ and $i_0 = \Delta y/2$. This corresponds to defining the following image coordinates as:

$$\begin{cases} x = j - j_0 = j - \frac{\Delta x}{2} \\ y = -(i - i_0) = \frac{\Delta y}{2} - i \end{cases} \quad (18)$$

When the detector circle angles are zero, its origin expressed in laboratory coordinates is:

$$O_D = \begin{bmatrix} d \\ 0 \\ 0 \end{bmatrix} \quad (19)$$

and its basis is:

$$D = [\widehat{\boldsymbol{d}}_x \quad \widehat{\boldsymbol{d}}_y \quad \widehat{\boldsymbol{d}}_z] = \begin{bmatrix} 0 & 0 & 0 \\ 0 & 1 & 0 \\ 0 & 0 & 1 \end{bmatrix} \quad (20)$$

When the detector circle angles $2\theta_h$, $2\theta_v$, $\omega$ are non-zero, the detector basis rotates according to:

$$D' = R_{xyz}D = [\widehat{\boldsymbol{d}}'_x \quad \widehat{\boldsymbol{d}}'_y \quad \widehat{\boldsymbol{d}}'_z] =$$

$$\begin{bmatrix} 0 & -\sin 2\theta_h \cos \omega - \cos 2\theta_h \sin 2\theta_v \sin \omega & \sin 2\theta_h \sin \omega - \cos 2\theta_h \sin 2\theta_v \cos \omega \\ 0 & \cos 2\theta_h \cos \omega - \sin 2\theta_h \sin 2\theta_v \sin \omega & -\cos 2\theta_2 \sin \omega - \sin 2\theta_h \sin 2\theta_v \cos \omega \\ 0 & \cos 2\theta_v \sin \omega & \cos 2\theta_v \cos \omega \end{bmatrix} \quad (21)$$

Therefore a ray intersecting the detector at a point $P'$ has the following coordinates in the detector reference system:

$$\begin{cases} x = P' \cdot \widehat{\boldsymbol{d}}'_y \\ y = P' \cdot \widehat{\boldsymbol{d}}'_z \end{cases} \quad (22)$$

which can also be rewritten as $[x \quad y \quad 0]^T = D'^T P'$. Eq. (22) can be rewritten by using the expression for $P'$ (eq. (16)) and the expression for $\widehat{d}'_y$ and $\widehat{d}'_{z2}$ (eq. (21)) in spherical coordinates

$$x = \frac{x_n}{\Delta}$$

$$y = \frac{y_n}{\Delta}$$

$$x_n = -d(\cos 2\Theta_{hL} \cos 2\Theta_{vL} \sin 2\theta_h \cos \omega - \sin 2\Theta_{vL} \cos 2\theta_v \sin \omega$$
$$- \sin 2\Theta_{hL} \cos 2\Theta_{vL} \cos 2\theta_h \cos \omega + \cos 2\Theta_{hL} \cos 2\Theta_{vL} \cos 2\theta_h \sin 2\theta_v \sin \omega \quad (23)$$
$$+ \sin 2\Theta_{hL} \cos 2\Theta_{vL} \sin 2\theta_h \sin 2\theta_v \sin \omega)$$

$$y_n = -d(\sin 2\Theta_{hL} \cos 2\Theta_{vL} \cos 2\theta_h \sin \omega - \sin 2\Theta_{vL} \cos 2\theta_v \cos \omega$$
$$- \cos 2\Theta_{hL} \cos 2\Theta_{vL} \sin 2\theta_h \sin \omega + \cos 2\Theta_{hL} \cos 2\Theta_{vL} \cos 2\theta_h \sin 2\theta_v \cos \omega$$
$$+ \sin 2\Theta_{hL} \cos 2\Theta_{vL} \sin 2\theta_h \sin 2\theta_v \cos \omega)$$

$$\Delta = \cos 2\theta_h \cos 2\theta_v \cos 2\Theta_v \cos 2\Theta_h + \sin 2\theta_h \cos 2\theta_v \cos 2\Theta_v \sin 2\Theta_h + \sin 2\theta_v \sin 2\Theta_v$$

and in Cartesian coordinates:

$$k_{fLx} = \frac{k_{fLxd}}{\Delta_k}$$

$$k_{fLy} = \frac{k_{fLyd}}{\Delta_k}$$

$$k_{fLxd} = -(x \sin 2\theta_h \cos \omega - d \cos 2\theta_h \cos 2\theta_v - y \sin 2\theta_h \sin \omega + y \cos 2\theta_h \sin 2\theta_v \cos \omega$$
$$+ x \cos 2\theta_h \sin 2\theta_v \sin \omega)$$

$$k_{fLyd} = -(y \cos 2\theta_h \sin \omega - d \sin 2\theta_h \cos 2\theta_v - x \cos 2\theta_h \cos \omega + y \sin 2\theta_h \sin 2\theta_v \cos \omega$$
$$+ x \sin 2\theta_h \sin 2\theta_v \sin \omega) \quad (24)$$

$$k_{fLz} = \sqrt{1 - k_{fLx}^2 - k_{fLy}^2}$$

$$\Delta_k = \sqrt{x^2 + y^2 + d^2}$$

$$2\Theta_{hL} = \operatorname{atan} \frac{k_{fLy}}{k_{fLx}}$$

$$2\Theta_{vL} = \operatorname{atan} \frac{k_{fLz}}{\sqrt{k_{fLx}^2 + k_{fLy}^2}}$$

The following calculations can be performed in MATLAB with Script 6, Supplementary Information.

**Case 1**. The detector is at its origin ($2\theta_h = 0, 2\theta_v = 0, \omega = 0$):

$$\hat{k}_{fL} = \frac{1}{\sqrt{x^2 + y^2 + d^2}} \begin{bmatrix} d \\ x \\ y \end{bmatrix} \quad (25)$$

$$2\Theta_{hL} = \text{atan2}(v_y, v_x) = \text{atan2}(x, d)$$

$$2\Theta_{vL} = \text{atan2}(y, \sqrt{d^2 + x^2})$$

**Case 2.** If we substitute $(x, y) = (0, 0)$ in eq. (25)

$$\hat{k}_{fL} = \begin{bmatrix} 1 \\ 0 \\ 0 \end{bmatrix} \quad (26)$$

and $2\Theta_{vL} = 0$ and $2\Theta_{hL} = 0$.

**Case 3.** With $\omega = 0$, if $2\Theta_h = 2\theta_h = 30°, 2\Theta_{vL} = 2\theta_{vL} = 30°$ ($\cos \pi/6 = \sqrt{3}/2$, $\sin \pi/6 = 1/2$), $(x, y) = (0,0)$, the projection of $\hat{k}_{fL}$ is at the origin of the detector (see Example 3 case 3 in the software).

### Rotating the Sample Circles

We now consider the last reference system, the sample coordinates system S, which is represented by the versors $\hat{s}_x, \hat{s}_y, \hat{s}_z$ (see **Figure 1d**). Following You's[30] formalism we indicate the crystal momentum transfer in the reciprocal space coordinates as:

$$\boldsymbol{h} = h\boldsymbol{b}_1 + k\boldsymbol{b}_2 + l\boldsymbol{b}_3 \quad (27)$$

where $\boldsymbol{b}_1, \boldsymbol{b}_2, \boldsymbol{b}_3$ are the reciprocal lattice vectors:[1,4,38,39] If the basis of the reciprocal space is different from the laboratory basis, we can then construct a matrix B so that the momentum transfer is expressed in the laboratory basis:[30]

$$\boldsymbol{h}_c = \text{B}\boldsymbol{h} \quad (28)$$

The crystal might have some intrinsic preferential orientation with respect to the sample substrate. This is represented by a rotation matrix U (orientation matrix) so that[30]:

$$\boldsymbol{h}_\phi = \text{U}\boldsymbol{h}_c \quad (29)$$

The momentum transfer $\boldsymbol{h}_\phi$ gives information about the crystal and its orientation. The diffraction condition is[30]:

$$\boldsymbol{h}_\phi = \boldsymbol{q}_S = \boldsymbol{k}_{fS} - \boldsymbol{k}_{iS} \quad (30)$$

where we have used the subscript $S$ to indicate that the vector $\boldsymbol{q}_S, \boldsymbol{k}_{fS}, \boldsymbol{k}_{iS}$ are expressed in the sample reference system.

Now we would like to see what happens to these vectors when the sample circle is rotated. We would like to calculate the effect of a sample rotation on the ray projected on the detector plane.

When all the sample circles are zero, the sample basis corresponds to the laboratory basis $S = L = [\hat{s}_x \quad \hat{s}_y \quad \hat{s}_z]$. The rotation of any of the sample circles reorients the sample basis and the sample or crystal itself, according to a rotation matrix given by the composition of these rotations. In case of sample rotating circles mounted one on each other, the way to build the total rotation matrix proceeds from the inner circle to the outer circle as $R = R_{out}R_{out-1} \cdots R_{in+1}R_{in}$. Here we only consider the case in which the pitch ($\alpha_i$) is rotated. Therefore, we will just use one rotation matrix $A_i$. As mentioned above, a positive pitch rotation corresponds to a clockwise rotation of the sample with respect to $\hat{l}_y$ (see **Figure 1d**), and the rotation matrix is:

$$A_i = \begin{bmatrix} \cos\alpha_i & 0 & -\sin\alpha_i \\ 0 & 1 & 0 \\ \sin\alpha_i & 0 & \cos\alpha_i \end{bmatrix} \tag{31}$$

A pitch rotation, changes the sample basis from $S = L = I$, where all the sample circles are zero, to:

$$S' = A_i L = A_i I = \begin{bmatrix} \cos\alpha_i & 0 & -\sin\alpha_i \\ 0 & 1 & 0 \\ \sin\alpha_i & 0 & \cos\alpha_i \end{bmatrix} \tag{32}$$

In this simple case, the total rotation matrix is simply $R = A_i$. In the presence of such a rotation the momentum transfer rotates according to[30]:

$$\boldsymbol{h}_R = A_i \boldsymbol{h}_\phi \tag{33}$$

The diffraction condition is[30]:

$$\begin{aligned} \boldsymbol{h}_R &= \boldsymbol{q}_L = \boldsymbol{k}_{fL} - \boldsymbol{k}_{iL} \\ \boldsymbol{q}_L &= A_i \boldsymbol{q}_S \\ \boldsymbol{k}_{fL} &= A_i \boldsymbol{k}_{fS} \\ \boldsymbol{k}_{iL} &= A_i \boldsymbol{k}_{iS} \end{aligned} \tag{34}$$

The matrix $A_i^{-1} = A^T$ rotates the wave vectors $\boldsymbol{k}_{fL}$ and $\boldsymbol{k}_{iL}$ counterclockwise about $\hat{l}_y$, and allows to express them as $\boldsymbol{k}_{fS}$ and $\boldsymbol{k}_{iS}$ in sample coordinates $\hat{s}_x, \hat{s}_y, \hat{s}_z$. Therefore the incident wave vectors in the laboratory and sample reference system are (see ref. [30] equation (12), where $\mathbf{Q}_L$ is our $\boldsymbol{q}_L$):

$$\boldsymbol{k}_{iL} = k \begin{bmatrix} 1 \\ 0 \\ 0 \end{bmatrix}$$

$$\boldsymbol{k}_{iS} = A_i^{-1} \boldsymbol{k}_{iL} = k \begin{bmatrix} \cos\alpha_i \\ 0 \\ -\sin\alpha \end{bmatrix} \tag{35}$$

The exit wave vectors in the laboratory and sample basis are:

$$\boldsymbol{k}_{fL} = k \begin{bmatrix} \cos 2\Theta_{vL} \cos 2\Theta_{hL} \\ \cos 2\Theta_{vL} \sin 2\Theta_{hL} \\ \sin 2\Theta_{vL} \end{bmatrix}$$

$$\boldsymbol{k}_{fS} = A_i^{-1} \boldsymbol{k}_{fL} = k \begin{bmatrix} \cos 2\Theta_{vS} \cos 2\Theta_{hS} \\ \cos 2\Theta_{vS} \sin 2\Theta_{hS} \\ \sin 2\Theta_{vS} \end{bmatrix} \quad (36)$$

where $2\Theta_{hS}$ and $\cos 2\Theta_{vS}$ are the azimuth and the elevation of $\hat{\boldsymbol{k}}_{fS} = \boldsymbol{k}_{fS}/k$ in spherical coordinates:

$$2\Theta_{hS} = \operatorname{atan2}(k_{fSy}, k_{fSx})$$
$$2\Theta_{vS} = \operatorname{atan2}(k_{fSz}, \sqrt{k_{fSx}^2 + k_{fSy}^2}) \quad (37)$$

The rotated scattering vector in the laboratory and sample reference system are:

$$\boldsymbol{q}_L = \boldsymbol{h}_R = \boldsymbol{k}_{fL} - \boldsymbol{k}_{iL} = k \begin{bmatrix} \cos 2\Theta_{vL} \cos 2\Theta_{hL} - 1 \\ \cos 2\Theta_{vL} \sin 2\Theta_{hL} \\ \sin 2\Theta_{vL} \end{bmatrix}$$

$$\boldsymbol{q}_S = \boldsymbol{h}_\phi = \boldsymbol{k}_{fS} - \boldsymbol{k}_{iS} = k \begin{bmatrix} \cos 2\Theta_{vS} \cos 2\Theta_{hS} - \cos \alpha_i \\ \cos 2\Theta_{vS} \sin 2\Theta_{hS} \\ \sin 2\Theta_{vS} + \sin \alpha_i \end{bmatrix} \quad (38)$$

If we convert the detector coordinates from pixels to $\boldsymbol{q}_S$ we will be able to observe the reflection $\boldsymbol{h}_\phi$ at the same point on the detector as $\alpha_i$ varies. This is the most important concept of this paragraph. If we exclude refraction effects[40], by employing $\boldsymbol{q}_S$ as the image coordinates we can remove distortion effects in the diffraction pattern introduced by a non-zero pitch. Note that eq. $\boldsymbol{q}_S$ is equivalent to the relations mapping the image of the untilted fibre from the detector plane into reciprocal space, which is reported in ref. [41].

Finally, the reciprocal space mapping conversion procedure can be summarized as follows:

1. Define $(x, y)$ pixel reference system for the diffraction image. There might be cases in which, when the detector and sample circle angles are zero, the direct beam is not at the centre of the detector. In this case the origin of the image $(x_0, y_0)$ cannot be at the centre of the diffraction pattern, and has to be set to the point where the direct beam hits the detector (eq. (18));
2. Extract image pixel coordinates $(x, y)$ from the diffraction image and retrieve $2\theta_h$, $2\theta_v$, $\omega$ and $d$. If the detector is mounted on the detector circles the distance $d$ can be calculated by tracking the direct beam, as shown in ref.[40]. The angles $2\theta_h$, $2\theta_v$, $\omega$ are immediately available if the detector is mounted on the detector circles. However there can be cases in which the detector is mounted elsewhere (e.g. linear stages). In this case a calibration material (e.g. silver behenate) can be used for the extraction of $2\theta_h$, $2\theta_v$, and $d$.
3. Use eq. (24) to convert $(x, y)$ to $\hat{\boldsymbol{k}}_{fL}$;
4. Use eq. (36) to obtain $\hat{\boldsymbol{k}}_{fS}$ from $\hat{\boldsymbol{k}}_{fL}$;
5. Use eq. (38) to calculate the momentum transfer or scattering vector $\boldsymbol{q}_S$ in the sample reference system;

6. Remap the intensities (interpolation) from the original diffraction image in $(x, y)$ coordinates into suitable 2D coordinates for $\boldsymbol{q}_S$, for example $q_{Sz}$ as the ordinate and $q_{Sxy} = \sqrt{q_{Sx}^2 + q_{Sy}^2}$ as the abscissa.

## Conclusion

In this work we have derived, in a rigorous way, the reciprocal space mapping equations for a '3D+1S' diffractometer in a way that is understandable to anyone with basic notions of linear algebra, geometry, and X-ray diffraction. With this set of equations, starting from the detector and sample circle angles and the distance sample-detector one can convert a diffraction image represented in pixel coordinates to the momentum transfer or scattering vector in the sample reference system.

## References

1. Birkholz, M. *Thin film analysis by X-ray scattering*. (John Wiley & Sons, 2006).
2. Vlieg, E., Van der Veen, J., Macdonald, J. & Miller, M. Angle calculations for a five-circle diffractometer used for surface X-ray diffraction. *Journal of applied crystallography* **20**, 330-337, (1987).
3. Feidenhans, R. Surface structure determination by X-ray diffraction. *Surface Science Reports* **10**, 105-188, (1989).
4. Als-Nielsen, J. & McMorrow, D. *Elements of modern X-ray physics*. (John Wiley & Sons, 2011).
5. Ungar, G., Liu, F., Zeng, X., Glettner, B., Prehm, M., Kieffer, R. *et al.* in *Journal of Physics: Conference Series.* 012032 (IOP Publishing).
6. Grelet, E., Dardel, S., Bock, H., Goldmann, M., Lacaze, E. & Nallet, F. Morphology of open films of discotic hexagonal columnar liquid crystals as probed by grazing incidence X-ray diffraction. *The European Physical Journal E* **31**, 343-349, (2010).
7. Perlich, J., Schwartzkopf, M., Körstgens, V., Erb, D., Risch, J., Müller-Buschbaum, P. *et al.* Pattern formation of colloidal suspensions by dip-coating: An in situ grazing incidence X-ray scattering study. *Physica status solidi (RRL)-Rapid Research Letters* **6**, 253-255, (2012).
8. Renaud, G., Lazzari, R. & Leroy, F. Probing surface and interface morphology with Grazing Incidence Small Angle X-Ray Scattering. *Surface Science Reports* **64**, 255-380, (2009).
9. Bikondoa, O., Carbone, D., Chamard, V. & Metzger, T. H. Ageing dynamics of ion bombardment induced self-organization processes. *Scientific Reports* **3**, 1850, (2013).
10. Joshi, G. R., Cooper, K., Lapinski, J., Engelberg, D. L., Bikondoa, O., Dowsett, M. G. *et al.* (NACE International).
11. Springell, R., Rennie, S., Costelle, L., Darnbrough, J., Stitt, C., Cocklin, E. *et al.* Water corrosion of spent nuclear fuel: radiolysis driven dissolution at the UO2/water interface. *Faraday Discussions* **180**, 301-311, (2015).
12. Katsouras, I., Asadi, K., Li, M., van Driel, T. B., Kjaer, K. S., Zhao, D. *et al.* The negative piezoelectric effect of the ferroelectric polymer poly(vinylidene fluoride). *Nat Mater*, (2015).
13. Müller-Buschbaum, P. Grazing incidence small-angle X-ray scattering: an advanced scattering technique for the investigation of nanostructured polymer films. *Anal Bioanal Chem* **376**, 3-10, (2003).
14. Dane, T. G., Cresswell, P. T., Pilkington, G. A., Lilliu, S., Macdonald, J. E., Prescott, S. W. *et al.* Oligo(aniline) nanofilms: from molecular architecture to microstructure. *Soft Matter* **9**, 10501-10511, (2013).
15. Al-Jawad, M., Steuwer, A., Kilcoyne, S. H., Shore, R. C., Cywinski, R. & Wood, D. J. 2D mapping of texture and lattice parameters of dental enamel. *Biomaterials* **28**, 2908-2914, (2007).
16. Simmons, L. M., Al-Jawad, M., Kilcoyne, S. H. & Wood, D. J. Distribution of enamel crystallite orientation through an entire tooth crown studied using synchrotron X-ray diffraction. *European journal of oral sciences* **119**, 19-24, (2011).

17	Beddoes, C. M., Case, C. P. & Briscoe, W. H. Understanding nanoparticle cellular entry: A physicochemical perspective. *Advances in Colloid and Interface Science* **218**, 48-68, (2015).
18	Renaud, G. Oxide surfaces and metal/oxide interfaces studied by grazing incidence X-ray scattering. *Surface Science Reports* **32**, 5-90, (1998).
19	Staniec, P. A., Parnell, A. J., Dunbar, A. D., Yi, H., Pearson, A. J., Wang, T. *et al.* The nanoscale morphology of a PCDTBT: PCBM photovoltaic blend. *Advanced Energy Materials* **1**, 499-504, (2011).
20	Agostinelli, T., Ferenczi, T. A., Pires, E., Foster, S., Maurano, A., Müller, C. *et al.* The role of alkane dithiols in controlling polymer crystallization in small band gap polymer: Fullerene solar cells. *Journal of Polymer Science Part B: Polymer Physics* **49**, 717-724, (2011).
21	Lilliu, S., Agostinelli, T., Hampton, M., Pires, E., Nelson, J. & Macdonald, J. E. The Influence of Substrate and Top Electrode on the Crystallization Dynamics of P3HT: PCBM Blends. *Energy Procedia* **31**, 60-68, (2012).
22	Lilliu, S., Alsari, M., Bikondoa, O., Macdonald, J. E. & Dahlem, M. S. Absence of Structural Impact of Noble Nanoparticles on P3HT:PCBM Blends for Plasmon-Enhanced Bulk-Heterojunction Organic Solar Cells Probed by Synchrotron GI-XRD. *Scientific Reports*, (2015).
23	Juraić, K., Gracin, D., Šantić, B., Meljanac, D., Zorić, N., Gajović, A. *et al.* GISAXS and GIWAXS analysis of amorphous–nanocrystalline silicon thin films. *Nuclear Instruments and Methods in Physics Research Section B: Beam Interactions with Materials and Atoms* **268**, 259-262, (2010).
24	Treat, N. D., Brady, M. A., Smith, G., Toney, M. F., Kramer, E. J., Hawker, C. J. *et al.* Interdiffusion of PCBM and P3HT reveals miscibility in a photovoltaically active blend. *Advanced Energy Materials* **1**, 82-89, (2011).
25	Rogers, J. T., Schmidt, K., Toney, M. F., Kramer, E. J. & Bazan, G. C. Structural order in bulk heterojunction films prepared with solvent additives. *Advanced Materials* **23**, 2284-2288, (2011).
26	Büchele, P., Richter, M., Tedde, S. F., Matt, G. J., Ankah, G. N., Fischer, R. *et al.* X-ray imaging with scintillator-sensitized hybrid organic photodetectors. *Nature Photonics*, (2015).
27	Tsao, H. N., Cho, D., Andreasen, J. W., Rouhanipour, A., Breiby, D. W., Pisula, W. *et al.* The Influence of Morphology on High-Performance Polymer Field-Effect Transistors. *Advanced Materials* **21**, 209-212, (2009).
28	Nelson, T. L., Young, T. M., Liu, J., Mishra, S. P., Belot, J. A., Balliet, C. L. *et al.* Transistor paint: high mobilities in small bandgap polymer semiconductor based on the strong acceptor, diketopyrrolopyrrole and strong donor, dithienopyrrole. *Advanced Materials* **22**, 4617-4621, (2010).
29	He, B. B. *Two-dimensional X-ray diffraction*. (John Wiley & Sons, 2011).
30	You, H. Angle calculations for a `4S+2D' six-circle diffractometer. *Journal of Applied Crystallography* **32**, 614-623, (1999).
31	Banchoff, T. & Wermer, J. *Linear algebra through geometry*. (Springer Science & Business Media, 2012).
32	Bloom, D. M. *Linear algebra and geometry*. (CUP Archive, 1979).
33	Kostrikin, A. I., Manin, Y. I. & Alferieff, M. E. *Linear algebra and geometry*. (Gordon and Breach Science Publishers, 1997).
34	Shafarevich, I. R. & Remizov, A. *Linear algebra and geometry*. (Springer Science & Business Media, 2012).
35	Hase, T. *Beamline Description*, <http://www2.warwick.ac.uk/fac/cross_fac/xmas/description/> (2015).
36	Boesecke, P. Reduction of two-dimensional small-and wide-angle X-ray scattering data. *Applied Crystallography*, (2007).
37	Murray, G. *Rotation About an Arbitrary Axis in 3 Dimensions*, <http://inside.mines.edu/fs_home/gmurray/ArbitraryAxisRotation/> (2013).

38  Woolfson, M. M. *An introduction to X-ray crystallography*. (Cambridge University Press, 1997).
39  Warren, B. E. *X-ray Diffraction*. (Courier Corporation, 1969).
40  Lilliu, S., Agostinelli, T., Pires, E., Hampton, M., Nelson, J. & Macdonald, J. E. Dynamics of crystallization and disorder during annealing of P3HT/PCBM bulk heterojunctions. *Macromolecules* **44**, 2725-2734, (2011).
41  Stribeck, N. & Nöchel, U. Direct mapping of fiber diffraction patterns into reciprocal space. *Journal of Applied Crystallography* **42**, 295-301, (2009).
42  Ohad, G. *fit_ellipse*, <http://www.mathworks.com/matlabcentral/fileexchange/3215-fit-ellipse> (2003).

# Supplementary Information

## Symbolic Calculations

The following scripts are independent from the *detector-GUI Matlab* software and require the Symbolic Math Toolbox.

**Script 1 – Arbitrary matrix rotation constructor**. This function is used in the following scripts.

```
function R = arb_rot_sim(a,vect)
% Rotation about arbitrary axis for symb calculations
%vect = vect / norm (vect);
u = vect(1); v = vect(2); w = vect(3);
u2 = u^2  ; v2 = v^2  ; w2 = w^2;
c = cos(a); s = sin(a);

R(1,1) = u2 + (1-u2)*c;
R(1,2) = u*v*(1-c) - w*s;
R(1,3) = u*w*(1-c) + v*s;
R(2,1) = u*v*(1-c) + w*s;
R(2,2) = v2 + (1-v2)*c;
R(2,3) = v*w*(1-c) - u*s;
R(3,1) = u*w*(1-c) - v*s;
R(3,2) = v*w*(1-c)+u*s;
R(3,3) = w2 + (1-w2)*c;
```

**Script 2 – Calculates rotation matrices**. See Tracking the Detector Movement.

```
%% Tracking the Detector Movement
clear all
syms h2 v2 om real  % detector rotation angles (azimuth, elevation, omega)
Rx = arb_rot_sim(om,[1 0 0]')  % eq.(2)
Ry = arb_rot_sim(v2,-[0 1 0]') % eq.(3)
Rz = arb_rot_sim(h2,[0 0 1]')  % eq.(4)
Rxyz = simplify(Rz*Ry*Rx)     % eq.(5), total rot. matrix
isequal(simplify(Rxyz'), simplify(Rxyz^-1)) % matrix orthogonality check
isequal(simplify(det(Rxyz)),1)       % det = 1 check
Ryz  = simplify(Rz*Ry)      % eq.(7), total rot. matrix with om = 0
```

**Script 3 – Projecting a ray onto the detector**. Running Script 2 is required before running this script. See Projecting the Exit Wave Vector onto the Detector.

```
%% Projecting a Ray onto the Detector
syms H2 V2 d real   % ray azimuth and elevation, sample-dectector distance
syms vx vy vz real  % ray Cartesian coordiantes
n = [1 0 0]'     % eq.(9), normal versor to unrotated detector
[V(1) V(2) V(3)] = sph2cart(H2, V2, 1); V = V' % eq.(12), ray sph. coord
n2  = Rxyz*n    % eq.(13), rotated versor
P = d*V/dot(V, n2)  % eq.(14), intersection point with detector
Pnum = d*V     % numerator in eq. (14)
Pden = dot(V, n2)   % numerator in eq. (14)

Vxyz = [vx vy vz]'      % eq.(15), ray Carthesian coordinates
P1 = d*Vxyz/dot(Vxyz, n2)   % eq.(15), intersection point with detector
P1num = d*Vxyz         % numerator in eq. (15)
P1den = dot(Vxyz, n2)     % numerator in eq. (15)
```

**Script 4 – Projecting a ray onto the detector.** Running Script 3 is required before running this script.

```
%% Ray Projection on Detector Coordinate System
dy = [0 1 0 ]'   % detector x axis
dz = [0 0 1 ]'   % detector y axis
dx = [0 0 0 ]'   % zero column, detector coord are in 2D
D = [dx , dy, dz] % eq.(18) detector basis
dy2 = Rxyz*dy   % eq.(19) roatated detector x axis
```

```
dz2 = Rxyz*dz    % eq.(19) rotated detector y axis
D2  = Rxyz*D     % eq.(19) rotated detector coordinate system

x = simplify(dot(P, dy2)); % eq.(21), point P in detector coordiantes
y = simplify(dot(P, dz2)); % eq.(21), point P in detector coordiantes

x1 = simplify(dot(P1, dy2)); % eq.(22), point P in detector coordiantes
y1 = simplify(dot(P1, dz2)); % eq.(22), point P in detector coordiantes
x1 = simplify(subs(x1, vz, sqrt(1-vx^2-vy^2))) % eq.(22)
y1 = simplify(subs(y1, vz, sqrt(1-vx^2-vy^2))) % eq.(22)

simplify(subs(x1, om, 0)) % eq.(24)
simplify(subs(y1, om, 0)) % eq.(24)
```

**Script 5 – Projecting a ray onto the detector.** Running Script 4 is required before running this script. See **Error! Reference source not found.**.

```
%% Finding a Ray from a Point on the Detector
syms X Y real
[vx_s vy_s] = solve([X == x1, Y == y1], vx, vy) % inverting eq.(22)
simplify(vx_s(2)) % eq.(25), vx (only the second solution makes sense)
simplify(vy_s(2)) % eq.(25), vy
vz_s = simplify(sqrt(1-vx_s(2)^2-vy_s(2)^2))
subs(simplify(vx_s(2)), om, 0) % eq.26, vx
subs(simplify(vy_s(2)), om, 0) % eq.26, vy
```

**Script 6 – Example 2.** Running script 5 is required before running this script. See **Error! Reference source not found.**.

```
%% Example 2
% case 1
subs(vx_s(2), [h2 v2 om], [0 0 0]) % eq.(40), vx
subs(vy_s(2), [h2 v2 om], [0 0 0]) % eq.(40), vy
simplify(subs(vz_s, [h2 v2 om], [0 0 0])) % eq.(4), vz

% case 3
c3x = double(subs(vx_s(2), [h2 v2 X Y d om], [pi/6 pi/6 0 0 1 0])) % eq.(43)
c3y = double(subs(vy_s(2), [h2 v2 X Y d om], [pi/6 pi/6 0 0 1 0])) % eq.(43)
c3z = double(simplify(subs(vz_s, [h2 v2 X Y d om], [pi/6 pi/6 0 0 1 0])))%
eq.(43)
c3 = [c3x c3y c3z];
[ca ce cr] = cart2sph(c3x, c3y, c3z);
ca_deg = rad2deg(ca) % eq.(44)
ce_deg = rad2deg(ce) % eq.(44)
```

**Script 7 – Scattering Vector.**

```
%% Defining the Ray as the Scattering Vector
syms lam real % wave vector amplitude 2pi/lambda
k = 2*pi/lam;
ki = [1 0 0]'
kf = V
q = kf-ki         % eq.(35)
simplify(acos(dot(ki, kf)))  % Bragg angle
simplify(sqrt(dot(q,q)))     % eq. (37) norm of q0

syms a b c real
a1 = [a 0 0]
a2 = [0 b 0]
a3 = [0 0 c]
b1 = 2*pi*(cross(a2, a3)/dot(a1,(cross(a2, a3))))
b2 = 2*pi*(cross(a3, a1)/dot(a2,(cross(a3, a1))))
b3 = 2*pi*(cross(a1, a2)/dot(a3,(cross(a1, a2))))
```

**Script 8 – Unrotated exit wave vector.** Running script 6 is required before running this script.

```
%% Rotating the sample circles
syms ai real
kfx = simplify(vx_s(2)) % eq.25, vx
kfy = simplify(vy_s(2)) % eq.25, vy
kfz = simplify(sqrt(1-vx_s(2)^2-vy_s(2)^2))
kf = [kfx; kfy; kfz]
Ai = arb_rot_sim(ai,-[0 1 0]'); % eq.(45) pitch rotation matrix
```